\RequirePackage{lineno}

\pdfoutput=1
\documentclass[twocolumn,superscriptaddress,floatfix,preprintnumbers,amssymb,amsmath]{revtex4}

\ifx\pdfoutput\undefined
  \usepackage[dvips]{graphicx}
\usepackage[dvips]{color}
\else
  \usepackage[pdftex]{graphicx}
\usepackage[pdftex]{color}
\fi

\usepackage{dcolumn}
\usepackage{latexsym}

\usepackage{units}
\usepackage{ulem}

\makeatother

\begin{document}
\title{DC measurement of dressed states in a coupled 100~GHz resonator
system using a single quasiparticle transistor as a sensitive microwave
detector}
\author{S.~V.~Lotkhov}
\affiliation{Physikalisch-Technische Bundesanstalt, Bundesallee 100, 38116 Braunschweig,
Germany}
\author{R.~Dolata}
\affiliation{Physikalisch-Technische Bundesanstalt, Bundesallee 100, 38116 Braunschweig,
Germany}
\author{M.~Khabipov}
\affiliation{Physikalisch-Technische Bundesanstalt, Bundesallee 100, 38116 Braunschweig,
Germany}

\begin{abstract}
We report on the on-chip detection of microwaves in the frequency
range around 100~GHz. For the purpose of detection, we employ a discrete
transport channel triggered in a superconducting single-electron transistor
by photon-assisted tunneling of quasiparticles. The technique is successfully applied
to observe the spectrum of the dressed states of a model cQED system consisting
of a superconducting coplanar resonator coupled to a quantum Josephson
oscillator. The dressed states appear as typical resonance anticrossing
exhibiting, in our case, an expectedly wide frequency splitting corresponding
to the Jaynes-Cummings coupling strength, $g/\pi\sim$~10~GHz. Due
to the high decay rate, $\gamma\sim$~20$\div$40~GHz, in the very
transparent Josephson junctions used, the strong coupling limit, $g\gg\gamma$,
which is required for qubit operation, is not achieved, and the photon population
in the resonator is low, $\langle n\rangle<$~1. Remarkably, the
continuous readout of the low population states demonstrates the high
microwave sensitivity of the detector.
\end{abstract}
\maketitle

Strong coupling between a microwave resonator and a quantum oscillator is
an important prerequisite for clear observation of circuit/cavity
quantum electrodynamics (cQED/CQED) effects \cite{Schoelkopf2008,Devoret2007}.
One of the significant advantages of superconducting circuit QED systems
over cavity QED with real atoms is a much stronger coupling between the
components. For example, for a Cooper pair box coupled to a coplanar waveguide (CPW)
resonator with a typical length $l$ of 10\ mm to 20~mm and mode frequency
$f_{{\rm r}}$ of 5~GHz to 10~GHz, the coupling strength $g$, which appears in the interaction term of
the Jaynes-Cummings Hamiltonian \cite{WallsMilburn}, routinely
achieves the level $g\sim2\pi\times$100~MHz (see, e.g., Refs.~\cite{Schoelkopf2008,Wallraff2004,Fink2008}).

Even stronger coupling should be expected in shorter resonators, $l\sim$~1~mm,
with the coupling strength, $\hbar g\propto e\sqrt{\hbar\omega_{{\rm r}}/2C}\propto l^{-1}$
\cite{Wallraff2004,Blais2004}, increasing linearly with the frequency,
$\omega_{{\rm r}}=2\pi f_{{\rm r}}=1/\sqrt{LC}\propto l^{-1}$ due
to the decrease in the effective inductance/capacitance $L,C\propto l$.
This could justify an interest in the compact quantum systems operating
in an upper microwave frequency range up to 100~GHz. On the other
hand, a substantial increase in the loss rate can critically impact
the quantum coherence as the result of a stronger capacitive coupling
to the electromagnetic environment at elevated frequencies and more
intensive generation of non-equilibrium quasiparticles (QP) (cf., e.g., Refs.~\cite{Barends2011,Gruenhaupt2018}).
For example, in a recent experiment \cite{Lehtinen2017}, a non-coherent
model approach was used successfully to describe an Aluminum SQUID
array up to the microwave frequencies $f$ from 50~GHz to 210~GHz,
exceeding the Cooper pair breaking energy, $2\Delta\approx$~400~$\mu$eV~$\approx$~$\,h\times$100~GHz.
An important basic question is therefore whether cQED behavior can be
observed at high microwave frequencies. However, the required measurement is
technically challenging, and dedicated microwave detection
techniques \cite{Lehtinen2017,BillangeonBasset,Lotkhov2012,Jalali2016,Lotkhov2016}
are necessary for a mK-setup of the dilution fridge at frequencies
beyond the standard scale, $f_{{\rm r}}\sim$~10~GHz.

\begin{figure*}[hbt]
\centering\includegraphics[width=1.8\columnwidth]{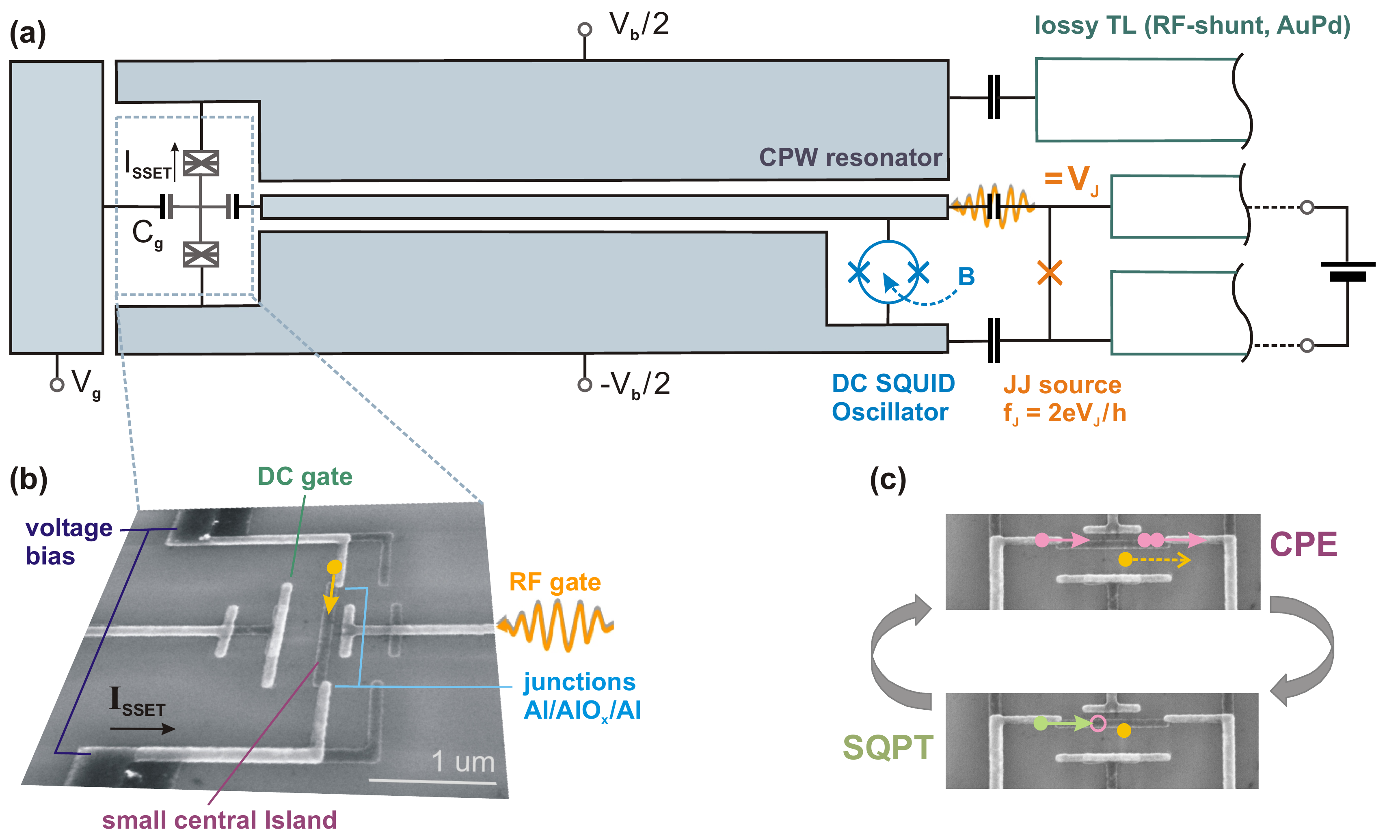}
\caption{(Color online) (a) Electrical circuit diagram of the experimental
device. The JJ source junction has an area $S\approx$~0.1~$\mu$m$^{2}$,
a critical current $I_{{\rm C}}\approx$~1.5~$\mu$A, and a Stewart-McCumber
parameter $\beta_{{\rm C}}\sim$~0.1, thus producing microwaves of
amplitude $V_{{\rm A}}\approx V_{{\rm C}}\approx$~75~$\mu$V, where
$V_{{\rm C}}=I_{{\rm C}}Z_{{\rm s}}$ is the Josephson characteristic
voltage. In the CPW, the central line (spacing) is 10~$\mu$m (5~$\mu$m)
(see text for more layout details). (b) Tilted SEM image of the SSET
detector and an example of Cooper pair splitting due to a PAT event. In
samples 1 and 2, $E_{{\rm C}}\approx$~150~$\mu$eV and the
sum tunnel resistance $R_{{\rm T}}\approx$~370~k$\Omega$. (c)
The transport cycle including consecutive SQPT and CPE cotunneling
steps in the presence of an unpaired QP. }
\label{Fig1}
\end{figure*}

In this Letter, we report our observations concerning the dressed states
of a CPW resonator directly coupled to a Josephson oscillator using the frequency
range $f\sim$~100~GHz. In order to access this range,
we implement an on-chip test bench, shown in Fig.~\ref{Fig1}(a),
based on Al/AlO$_{{\rm x}}$/Al-junctions and fully controlled via dc signals from
room temperature electronics. As a microwave source, we use an overdamped
Josephson junction (JJ source), and for the photon detection, we
utilize a quasiparticle sensing regime based on a photon-assisted
tunneling (PAT) effect in a superconducting single-electron transistor
(SSET), shown in Fig.~\ref{Fig1}(b) (cf. Refs.~\cite{Jalali2016,Lotkhov2016}).
A significant gain in microwave sensitivity appears in our detection
circuit owing to an intrinsic photon-electron multiplying mechanism
described below.

The experimental circuit was fabricated using the shadow evaporation
technique \cite{DolanAPL1977} and studied in a shielded DC setup
at $T=15$~mK with an integration constant, $\tau_{{\rm int}}\sim$~0.5~s,
that was sufficiently long for measuring the sub-pA currents.
All superconducting components
were integrated into a single evaporation mask and included three
successive layers of aluminum. Mild (10 min at oxygen pressure 1~Pa)
and heavy (15~min, 25~Pa) oxidation steps were performed for the
first and the second Al layers, respectively, to form the tunnel junctions
for the JJ source/Josephson oscillator and for the high-ohmic SSET
device. As a resistive shunt attached to the JJ source (see the right hand part
of Fig.~\ref{Fig1}(a)), we used a finite-loss AuPd coplanar transmission line (TL)
with a specific high-frequency impedance of $Z_{{\rm s}}\approx$~50~$\Omega$.
The line was 2.3~mm long and it was terminated at the opposite end by means
of a 6~$\Omega$ section of the AuPd film used as a cold part of the DC biasing
circuitry. The full DC load resistance seen by the Josephson junction
was $R_{{\rm L}}\approx$~25~$\Omega$, which was low enough for stabilization
of the Josephson voltage, $V_{{\rm J}}=hf_{{\rm J}}/2e$, and the base
frequency $f_{{\rm J}}$ of the Josephson generation (cf. Ref.~\cite{Jalali2014condmatt}).

The detection principle is based on the remarkably high quasiparticle
sensitivity of the tunnel current $I_{{\rm SSET}}$ at low bias voltages,
$V_{{\rm b}} < 4\Delta/e$, where single-electron tunneling is
suppressed. As discussed in Refs.~\cite{Brink1991,Hadley1998}, in
the bias range of 200\ $\mu V\leq V_{{\rm b}}\leq$~400~$\mu$V,
a non-vanishing current appears as a time-correlated sequence of single quasiparticle
tunneling (SQPT) and Cooper pair - Electron (CPE) cotunneling transfers
(see the cycle depicted in Fig.~\ref{Fig1}(c)). The SQPT/CPE sequence
persists provided there is at least one unpaired QP hosted by the
central island. It starts with a low-rate SQPT process (see the estimate below) transferring an electron into
the island and pairing it with the odd QP, followed by CPE cotunneling, a much faster step discharging the island,
but leaving behind another unpaired QP that will be incorporated into the next cycle.
The density of QPs (and thus the rate of SQPT) depends crucially both on the parity effect
(see, e.g., Ref.~\cite{Parity}) and on the intensity of the incident or
ambient microwaves (see, e.g., Refs.~\cite{Barends2011,Pekola2010,Kemppinen2011,SairaPRB2012}).
The microwave photon energy $E_{{\rm ph}}$ must exceed the activation
threshold for photon-assisted single-electron tunneling,
$E_{{\rm ph}}\agt E_{{\rm PAT}}=2\Delta+E_{{\rm C}}(1\mp2n_{{\rm g}}+2m)-eV_{{\rm b}}/2$.
Here, $n_{{\rm g}}\equiv C_{{\rm g}}V_{{\rm g}}/e$ is the gate charge,
$C_{{\rm g}}$ ($V_{{\rm g}}$) denotes the gate capacitance (voltage),
$E_{{\rm C}}\equiv e^{2}/2C_{\Sigma}$ is the charging energy, $C_{\Sigma}$ is
the total capacitance of the central island, and $m$ is an even integer
number.

\begin{figure}[hpt]
\centering\includegraphics[width=1\columnwidth]{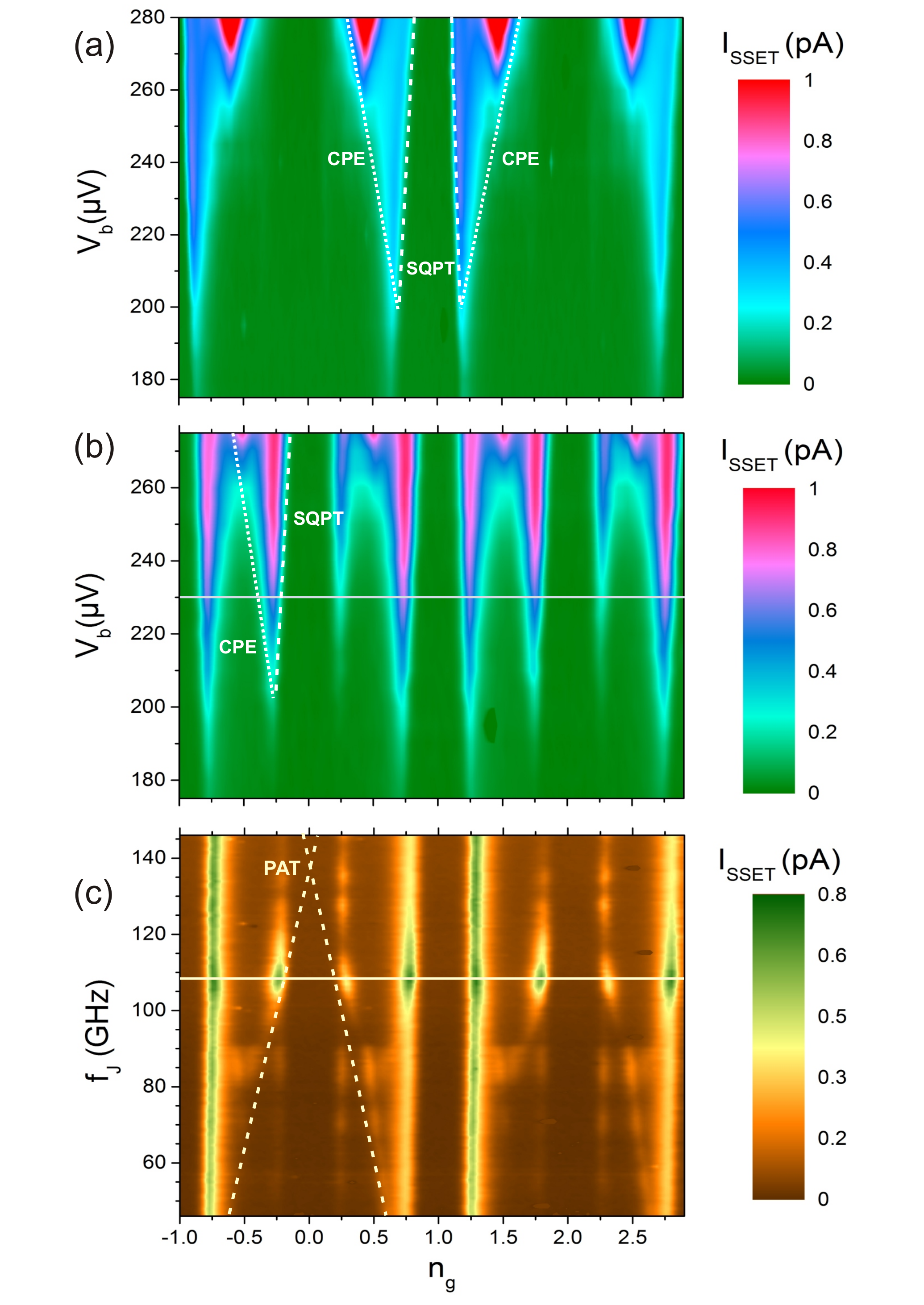}
\caption{(Color online) Detector current $I_{{\rm SSET}}$ measured on a test sample without
qubit, with a weaker JJ source: $I_{{\rm C}}\approx$~330~nA, $V_{{\rm C}}\approx$~13~$\mu$V,
and deviating SSET parameters: $E_{{\rm C}}\approx$~250~$\mu$eV,
and $R_{{\rm T}}\approx$~185~k$\Omega$. The diagrams show the
current pattern (a) without irradiation and (b) under irradiation
from the JJ source tuned to the resonant peak of CPW {[}the solid
line in (c){]}. The dotted (dashed) lines indicate the thresholds
for CPE cotunneling (SQPT) processes described in the text. (c) Transmission
spectrum of the CPW resonator collected via changing the Josephson
voltage (frequency) $V_{{\rm J}}$($f_{{\rm J}}$) in the JJ source
and measuring the signal of the detector biased at voltage $V_{{\rm b}}=$~230~$\mu$V {[}the solid
line in (b){]}. The dashed lines are the thresholds of the PAT processes
of the kind depicted in Fig.~\ref{Fig1}(b). }
\label{Fig2}
\end{figure}

Further details of the detector operation can be illustrated by using the data in Fig.~\ref{Fig2},
which was obtained for a test sample without a quantum oscillator. Without irradiation,
see Fig.~\ref{Fig2}(a), a clear sub-pA current pattern appears
in the odd gate domains, $m+1/2\le n_{{\rm g}}\le m+3/2$, with one unpaired QP residing
in the island most of the time. By contrast, the current landscape
is almost suppressed in the even domains, $m-1/2\le n_{{\rm g}}\le m+1/2$,
with all QPs being paired. The current appears at the voltage thresholds
$V_{{\rm b}}\agt2E_{{\rm C}}(\pm2n_{{\rm g}}-1+2m)/e$, for SQPT,
and $V_{{\rm b}}\agt\left[4\Delta+2E_{{\rm C}}(1\mp2n_{{\rm g}}+2m)\right]/3e$,
for the CPE cotunneling. The expectation time $\tau_{{\rm CPE}}$ of the CPE cotunneling greatly exceeds
the time $\tau_{{\rm SQPT}}$ for SQPT. Indeed, the CPE/CPE cotunneling
current, which is onset in the triangles (color online: red) at the
top of the diagram in Fig.~\ref{Fig2}(a) (see also Ref.~\cite{Hadley1998}) and reaches the values
$I_{{\rm SSET}}\sim$~10~pA beyond the diagram scope, provides an
estimate $\tau_{{\rm CPE}}\sim$~25~ns. On the other hand, a much
lower CPE/SQPT current measured, $I_{{\rm SSET}}\approx$~0.2-0.5~pA, corresponds
to the cycle period on the scale of a microsecond. This period is
obviously limited by the rate of SQPT, $\Gamma_{{\rm SQPT}}^{-1}=\tau_{{\rm SQPT}}\approx$~0.7-1.6~$\mu$s.

The high $\tau_{{\rm SQPT}}/\tau_{{\rm CPE}}$
ratio provides the conditions for an intrinsic mechanism of current
triggering due to photon absorption. In charge dynamics, the PAT process
shown in Fig.~\ref{Fig1}(b) is followed directly by the charge relaxation
act via CPE cotunneling, shown in Fig.~\ref{Fig1}(c). The resulting
state with two unpaired QPs in the island invokes the cyclic SQPT/CPE
transport similar to that in the autonomous mode. Accordingly, the
tunnel current should consist of the trains of CPE/SQPT pulses, persisting
until the unpaired QP escapes from the island in advance of the CPE
event. The most probable escape process, shown by the dashed arrow
in Fig.~\ref{Fig1}(c), has the rate close to $\Gamma_{{\rm SQPT}}$,
competing with the rate of CPE cotunneling, $\tau_{{\rm CPE}}^{-1}$.
The number of cycles in the train $N_{{\rm p}}$ is thus defined by
the times ratio, $N_{{\rm p}}\sim\tau_{{\rm SQPT}}/\tau_{{\rm CPE}}\sim$~50,
and the train duration is an $N_{{\rm p}}$-multiple of the cycle
time, $\tau_{{\rm QP}}\sim\tau_{{\rm SQPT}}\times\tau_{{\rm SQPT}}/\tau_{{\rm CPE}}\sim$~60~$\mu$s.
A more accurate description of QP dynamics (which is beyond the scope of this
Letter) should include a QP recombination rate in the island and requires
the master equation approach \cite{MaisiPRL2013}.

As expected, the detector signal is found to increase significantly
(see Fig.~\ref{Fig2}(b)), due to microwave irradiation arriving at
the detector via the CPW resonator. The current diagram appears to
be almost 1e-periodic, by exhibiting similar currents in the odd and
even domains, which indicates a QP population about unity in the
island. The signal dependence on the Josephson frequency (i.e., microwave
photon energy), which is shown in Fig.~\ref{Fig2}(c), confirms a clear resonant
peak structure at $f_{{\rm J}}\approx$~108~GHz, but also reveals
an inferior pattern of stray box resonances and higher Josephson harmonics.
Similar to Refs.~\cite{Jalali2016,Lotkhov2016}, we roughly estimated the
rate of photon-assisted tunneling in SSET irradiated under the conditions of
the resonant peak and the gate tuned to the sensitive point, $n_{{\rm g}}\approx -0.25$.
The obtained value, $\Gamma_{{\rm ph}}\sim10^{4}$~s$^{-1}$,
corresponds to an extremely low level of energy dissipation from
the CPW into the detector, $W_{{\rm D}}=\Gamma_{{\rm ph}}\times hf_{{\rm J}}\sim$~0.7~aW~$\approx$~-151~dBm.
Furthermore, relating $\Gamma_{{\rm ph}}$ to the signal peak value,
$I_{{\rm SSET}}\approx$~0.6~pA $\sim4\times10^{6}$~electrons
per second, it was possible to obtain a reasonable estimate, $N_{{\rm p}}\sim200$ cycles per
photon.

The quantum oscillator coupled to the CPW resonator was designed as
a Josephson DC SQUID with a loop of area $A\approx~40~\mu m^{2}$
and two highly transparent, $0.25~\mu m^{2}$ Josephson tunnel
junctions with a total critical current $I_{{\rm C}}(B=0)=I_{{\rm C1}}+I_{{\rm C2}}\approx$~8~$\mu$A
and a sum tunnel capacitance $C_{{\rm J}}\approx$~22~fF (a fitted
value, see below). As shown in Fig.~\ref{Fig3}(a), the magnetic
field $B$ applied to the SQUID loop periodically modulates the microwave
signal transmitted to the detector. The measured period, $\Delta B\approx~14~\mu T<\Phi_{{\rm 0}}/A\approx50~\mu T$,
is smaller than that expected for the loop area presumably due to a
flux concentration effect in the slots of the CPW line. On the
other hand, the actual flux and field values, $\Phi=\tilde{B}\times A$,
can be calibrated directly using the plot periodicity, $\Delta\Phi=\Phi_{{\rm 0}}$.
The plasma frequency of the SQUID, $\omega_{{\rm p}}(\Phi)=\sqrt{2eI_{{\rm C}}(\Phi)/\hbar C_{{\rm J}}}$,
is estimated to vary in a wide range up to $\omega_{{\rm p}}(0)\sim2\pi\times$170~GHz
and the Josephson-to-charging energy ratio approaches the values,
$E_{{\rm J}}/E_{{\rm CJ}}\sim$~4000, well within the limit of the transmon-type
qubits \cite{Koch2007}.

Due to the high detector sensitivity, we succeeded in observing
the dressed states \cite{Wallraff2004,Blais2004} in the superconducting
resonator system as a well-pronounced anticrossing, see Fig.~\ref{Fig3}(b),
between the microwave and qubit resonances at low detuning, $\delta=\lvert\omega_{{\rm p}}-\omega_{{\rm r}}\rvert\ll g$.
The uncoupled resonant frequencies $\omega_{{\rm r}}$ and $\omega_{{\rm p}}$
vary along the dashed lines and, on the state diagram in Fig.~\ref{Fig3}(c),
correspond to the resonant transitions from the ground state $|g,0\rangle$
to the single-photon state $|g,1\rangle$ or to the lowest excited
state of the qubit $|e,0\rangle$, respectively. The symmetric and
antisymmetric dressed states appear for the coupled system as the
coherent superpositions, $|\pm\rangle\approx1/\sqrt{2}(|g,1\rangle\pm|e,0\rangle)$
($1/\sqrt{2}$ is an exact prefactor at $\delta=$~0) and give rise
to resonances at \cite{Blais2004}:
\begin{equation}
\omega_{\pm}(\Phi)=\frac{\omega_{{\rm p}}(\Phi)+\omega_{{\rm r}}\pm\sqrt{4g^{2}+\left[\omega_{{\rm p}}(\Phi)-\omega_{{\rm r}}\right]^{2}}}{2},\label{pmstates}
\end{equation}
shown by the solid lines in Fig.~\ref{Fig3}(b). The lines are fitted
to the plot by adjusting the values of $C_{{\rm J}}$ and the SQUID
asymmetry factor, $d=\lvert I_{{\rm C1}}-I_{{\rm C2}}\rvert/(I_{{\rm C1}}+I_{{\rm C2}})\approx$~0.16.
The frequency splitting interval, equal to $g/\pi=e\sqrt{\omega_{{\rm r}}/hC}\approx$~9.2~GHz,
was calculated directly, based on the resonator geometry \cite{GoepplJAP2008-1}.
The signal peaks, observed at $\Phi\approx0.42\,\Phi_{{\rm 0}}$ and
$\approx0.58\thinspace\Phi_{{\rm 0}}$, appear in the vicinity of the
zero detuning points, presumably due to the resonant increase in the
qubit impedance, causing more efficient matching to the microwave source.
Finally, we note that the anticrossing pattern was studied for two
different CPW resonators with the microwave parameters summarized
in Table~\ref{Sampa}.

\begin{figure}[tb]
\centering\includegraphics[width=1.05\columnwidth]{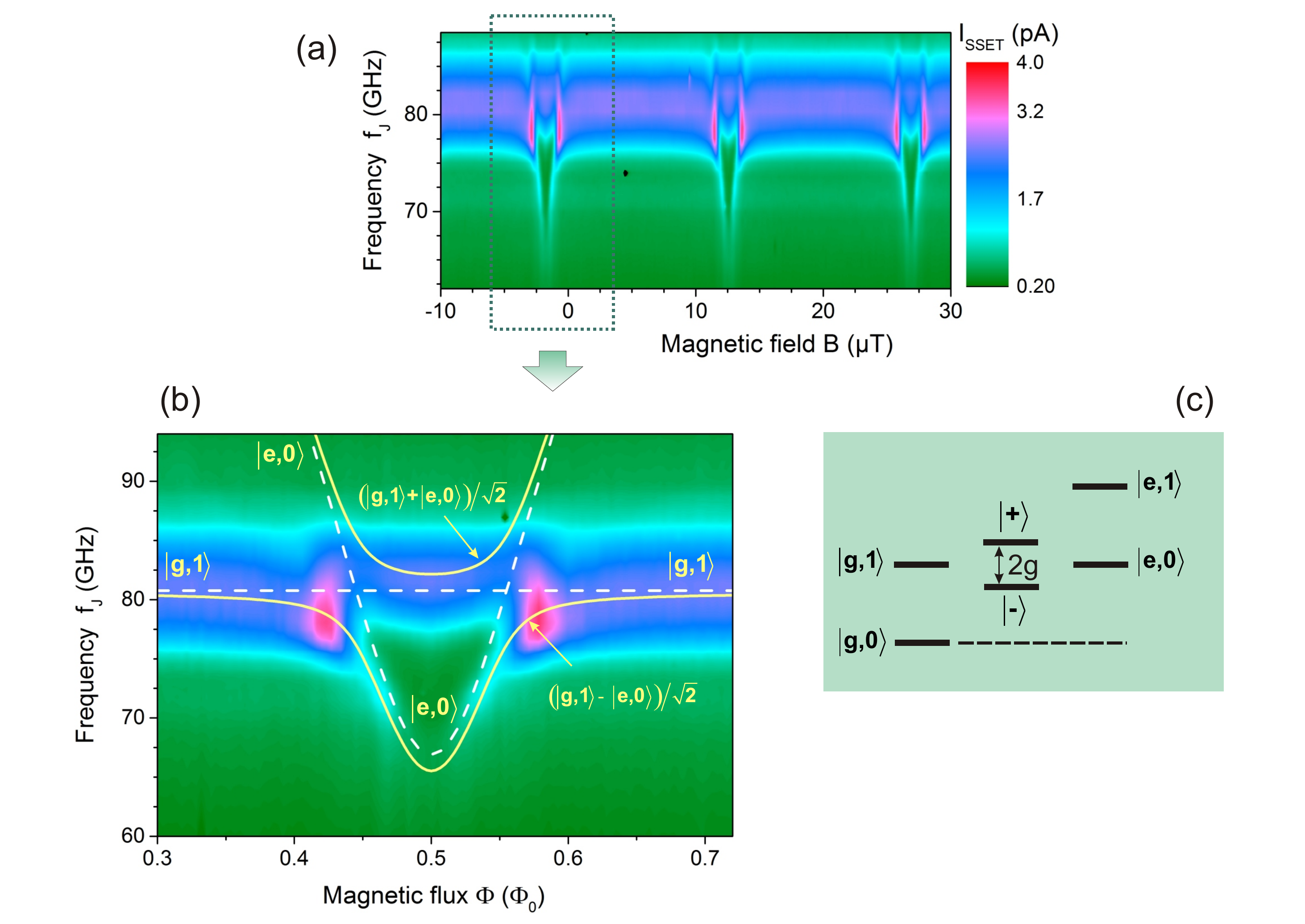}
\caption{(Color online) Dressed state characterization on the sample 1: (a)
The signal $I_{{\rm SSET}}$ vs. excitation frequency $f_{{\rm J}}$
and the magnetic field $B$. The bias and gate voltage settings of
the detector were chosen in the most sensitive point in respect to
the microwave signal. (b) Blow up diagram for the plot in (a): a resonance
anticrossing as a function of magnetic flux in units of flux quantum
$\Phi_{{\rm 0}}=h/2e=2.07\times10^{-15}$~Wb. See text for further
discussion. (c) State diagram for the lowest uncoupled ($|g,0\rangle$,
$|g,1\rangle$, $|e,0\rangle$ and $|e,1\rangle$) and dressed $|\pm\rangle$
atom-photon states at zero detuning \cite{Blais2004}. }
\label{Fig3}
\end{figure}
\begin{table}[tbh]
\caption{Sample parameters.}
\label{Sampa} \begin{ruledtabular} %
\begin{tabular}{ccccccl}
No.  & $l$,~mm & $f_{\text{r}}$,~GHz  & $g/2\pi$,~GHz  & $\Delta f$,~GHz  & $\gamma$,~GHz  & $\langle n\rangle$\tabularnewline
\hline
1  & 0.68  & 81  & 4.6  & 7.2  & 43  & 0.2 \\
2  & 0.8  & 70  & 3.9  & 3.8  & 21  & 0.6 \\
\end{tabular}\end{ruledtabular}
\end{table}
As compared to the unloaded test samples, with the detected linewidth
of the CPW resonances, $\Delta f\approx$~1.7~GHz, and the Josephson
generation linewidth of the source, $\Delta f_{{\rm J}}\approx$~0.4~GHz
\cite{LikharevBlueBook}, a significant broadening of the resonance
is observed for the samples including the SQUID junctions. We explain
this, on a qualitative level, as being the result of a strong quasiparticle
subgap leakage due to the high junction transparency (see, e.g., Refs.~\cite{LotkhovSIS2006,Sandberg2008})
on the one hand, and the PAT effect \cite{TienGordon1963} on the
other hand. The higher-frequency peak exhibits a stronger broadening,
thus emphasizing the significance for the leakage current of the photon
energy approaching the Cooper pair breaking threshold, $2\Delta/h\approx$~100~GHz.

Under continuous irradiation, the total loss rate in the coupled
CPW-plus-oscillator system, $\gamma=2\pi(\Delta f-\Delta f_{{\rm J}})$,
is counteracted by the power input from the JJ source, $P_{{\rm in}}\sim$~0.1~pW.
Due to the intensive losses, the average number $\langle n\rangle$
of photons in our CPW resonator, $\langle n\rangle=P_{{\rm in}}/(\gamma\hbar\omega_{{\rm r}})$,
is estimated to be lower than unity (see Table~\ref{Sampa}). On
the one hand, this may be a detector's figure of merit, for its
clear signal indicates the proper sensitivity of the resonator readout.
On the other hand, since the decay rate $\gamma$ exceeds Rabi frequency
$g/2\pi$, as both are in the GHz-range, no Rabi oscillation can exist
under the present conditions. Much lower dissipation (and much more
pronounced anharmonicity) can be expected for a SQUID oscillator with
sub-100~nm junctions of the same high barrier transparency (and thus
of the frequency $\omega_{{\rm r}}>$~100~GHz). Furthermore, a more
detailed study is envisaged of the loss rate in a wide range of resonator
frequencies.

To conclude, on the basis of a superconducting single-electron transistor,
we have developed a highly sensitive on-chip detection technique for
microwave frequencies on the scale of 100~GHz. An important element
of this technique is a mechanism of photon-activated current triggering
that facilitates a batch electron transfer for each absorbed photon.
The detection technique was used to observe the two lowest dressed states
in a quantum system with a CPW resonator coupled to a Josephson junction
oscillator. This observation demonstrated a detector sensitivity down
to very low photon populations in the resonator, $\langle n\rangle<$~1.
Further applications of the detector for on-chip studies of mesoscopic
devices are in progress.

We would like to acknowledge the technical support of V.~Rogalya and T.~Weimann.
This work was funded in part by the PARAWAVE Joint Research Project.
This project has received funding from the EMPIR program co-financed
by the Participating States and from the European Union\textquoteright s
Horizon 2020 research and innovation program. The measurement data
for this paper is available at: \url{https://doi.org/10.7795/720.20190617}
(Lotkhov et al., 2019).

\end{document}